\documentclass{iopart}

\usepackage{iopams,graphicx,hyperref}  
\begin{document}

\title{Universal Behavior of Entanglement in 2D Quantum Critical Dimer Models}

\author{Benjamin Hsu and Eduardo Fradkin}

\address{Department of Physics, University of Illinois at Urbana-Champaign, 1110 West Green Street, Urbana, Illinois 61801-3080, USA}
\eads{\mailto{bhsu2@illinois.edu}, \mailto{efradkin@illinois.edu} }

\begin{abstract}
We examine the scaling behavior of the entanglement entropy for the 2D quantum dimer model (QDM) at criticality and derive
 the universal finite sub-leading correction $\gamma_{QCP}$.  We compute the value of $\gamma_{QCP}$ without approximation working directly with the wave function of a generalized 2D QDM at the Rokhsar-Kivelson QCP in 
the continuum limit. Using the replica approach, we construct the conformal boundary state corresponding to the cyclic identification of $n$-copies along the 
boundary of the observed region. We find that the universal finite term is $\gamma_{QCP}=\ln R-1/2$ where $R$ is the compactification radius of the bose field theory quantum Lifshitz model, the effective 
field theory of the 2D QDM at quantum criticality.  We also demonstrated that the entanglement spectrum of the critical wave function on a large but finite region is described by the characters of the underlying 
conformal field theory. It is shown that this is formally related to the problems of quantum 
Brownian motion on $n$-dimensional lattices or equivalently a system of strings interacting with a brane containing a background electromagnetic field and can be written as an expectation value of a vertex operator.
\end{abstract}

\date{\today}
\maketitle

\section{Introduction}

Entanglement is one of the strangest features of quantum mechanics. Few would disagree that it distinguishes quantum mechanics unequivocally from classical 
physics. While it has played an essential role in the understanding of quantum mechanics, paradoxically it has been mostly absent from the theory of quantum phase 
transitions. As in the case of classical critical phenomena, the theory of a quantum phase transition is naturally based on the description of the 
scaling behavior of local  observables. 
Indeed, much of the theory of quantum criticality is based on this extension of the framework of classical criticality allowing for the natural dynamics specified by the 
quantum mechanical Hamiltonian of the system.\cite{sachdev-book} In this standard description the main effect of quantum mechanics is captured by the dynamic 
critical exponent associated $z$ that governs the relative scaling of space and time and is determined by the dynamics of the quantum system.
It is a major theoretical challenge is to understand the role of observables that have no classical analogue in quantum criticality. Quantum entanglement 
has been proposed as a candidate for such a measure.\cite{amico-2008}

While there are other measures of entanglement, the entanglement entropy has proven to be the most tractable analytically and simple to define. 
One begins with a pure state of a bipartite quantum system $A\cup B$. By restricting oneself to observing either $A$ or $B$ subregion, the subsystem is described by  
a mixed state with a non-trivial reduced density matrix. Suppose that $A$ is the observed region so that the degrees of freedom in $B$ are traced out. The reduced 
density matrix is $\rho_{A}=\textrm{Tr}_{B} \,\rho_{A\cup B} $. Non-local quantum correlations are then encoded in the von Neumann entropy, 
$S=-\textrm{Tr} \, \rho_{A} \ln \rho_{A}$. For the case of a total system $A \cup B$ in a system in a pure state, the entanglement entropy is symmetric  $S_{A} = S_{B}$ 
so that $S$ should only depend on common properties of the two regions. Given the non-local character of the entanglement entropy, its behavior in extended 
systems and quantum fields theories has a strong geometric flavor. An important early was the realization that in systems away from quantum 
criticality in $D$ space dimensions, systems dominated by short distance physics, the von Neumann entanglement entropy scales as the {\em area} of the observed 
region, $S \sim \mu \ell^{D-1}+\ldots$, where $\ell$ is the linear size of the region (say $A$) and $\mu$ is a non-universal constant.\cite{Srednicki1993,Bombelli1986}

The scaling behavior of the entanglement entropy  has been studied extensively in 1D quantum critical systems and it is by now reasonably well understood. 
Such systems are effectively relativistic ($z=1$) $1+1$ dimensional conformal field theories (CFT).
It has been shown that in 1D the entanglement entropy of a subsystem $A$ of linear size $\ell$  of an otherwise infinite system ({\it i.e.\/} of linear size $L \to \infty$) 
obeys a logarithmic scaling law,\cite{Callan1994,Holzhey1994,Calabrese2004,Vidal2003} $S\sim \frac{c}{3} \ln (\frac{\ell}{a})+\ldots$, where $c$ 
is the central charge of the CFT, and $a$ is the short distance cutoff. The growing popularity of entanglement entropy study is due to the fact that many 
universal properties of quantum systems like the central charge, excitation spectra or boundary entropy can be extracted from the entanglement entropy without the 
need to specify an observable: all that is needed is a consistent partition of the system. In addition,  the \emph{dynamical} entanglement 
entropy, entanglement generated in a quantum quench, has also been studied recently for a number of one-dimensional quantum critical 
systems.\cite{Hsu2009b,Lehur2009,KlichLevitov,CalabreseCardy-CFT} Even in strongly disordered quantum systems, which do not have a local order parameter such as random spin chains at infinite disorder fixed points\cite{refael-2009} 
as well as at the Anderson transition and quantum Hall plateau transition,\cite{Chakravarty}  the scaling behavior of the entanglement entropy has been shown to be  
a faithful measure of quantum criticality.

However, much less is known about the scaling of entanglement in spatial dimensions $D>1$, and how it relates to the scaling properties of local observable. While the 
leading scaling behavior of  the entanglement entropy is the area law, its prefactor is not universal, except in $D=1$ where the area law becomes the universal 
logarithmic scaling. For $D>1$ the only universal contributions may only arise from sub-leading terms (relative to the area law).
 The study of these universal corrections requires more detailed and subtle calculations. In general the 
situation is not as clear. An exception is the special case of {\em topological phases} in $D=2$, where there is a finite, i.e. $O(\ell^0)$,
universal correction to the area law which is given in terms of the topological invariants of the 
effective topological field theory of the phase. \cite{Kitaev2006,Levin2006,Dong2008}
Recent work has showed that several quantum critical systems in $D=2$ do not have logarithmic sub-leading corrections and, instead, have a {\em universal} finite term.\cite{Fradkin2006,Hsu2009,pasquier2009,Metlitski,FradkinReview}

In this paper we reexamine the scaling behavior of entanglement in a special class of quantum (multi)critical points in $D=2$ with dynamic scaling exponent $z=2$. 
These conformal quantum  critical points have the special property that the amplitudes of field configurations $\{ |\phi\rangle \}$ in their wave functions have local 
scale-invariant weights.\cite{Ardonne2004} Simple examples of such systems are 2D quantum dimer models\cite{Rokhsar1988} 
and their generalizations.\cite{Ardonne2004,Castelnovo2005,Fendley2006,Fendley2008} The norm of these ground state wavefunctions is thus equivalent to a partition function of a suitable two 
dimensional classical statistical model at criticality. 
Labeling a configuration of the classical statistical model by the field $\phi$, this can be expressed,
\begin{equation}
	\vert\vert \Psi_{0} \vert\vert^{2} = \int D\phi \, e^{-S(\phi)}. \label{eqn:gs}
\end{equation}
With the explicit form of the wavefunction, one can compute the entanglement entropy by constructing the replicated reduced density matrix, 
$\textrm{Tr }\rho_{A}^{n}$.  For these models it was shown that the entanglement entropy has a universal sub-leading correction, \cite{Hsu2009}
 \begin{equation}
 S_{QCP}=\mu \ell + \gamma_{QCP}+\ldots \, ,
 \label{eq:SQCP}
 \end{equation}
 where $\gamma_{QCP}=\ln R$ for the case of the quantum dimer models at their quantum critical (Rokhsar-Kivelson) point. Here $R$ is the compactification radius 
 (see below) of the coarse-grained height model, dual to the QDM, the quantum Lifshitz model.\cite{Ardonne2004,Moessner2002,Henley1997}
 The same scaling behavior has been shown to hold, within the $\epsilon$-expansion,  in relativistic $\phi^4$ quantum field theory, 
 the prototype of a quantum phase transition.\cite{Metlitski}

 In Ref.\cite{Fradkin2006} it was shown that the  computation of  the ``spectral moment'' of the reduced density matrix $S_n=\textrm{tr} \rho_A^n$ for the case of local 
 scale-invariant wave functions is the same as the computation  of a ratio in which the numerator is given by a classical partition function of the form of 
 Eq.\eref{eqn:gs} for $n$ copies of the system which are required to agree on the boundary of region $A$, while in the denominator no such constraint is imposed. 
 Posed in this way, the computation of $S_n$, and the von Neumann entanglement entropy $S=\lim_{n \to 1} \frac{1}{1-n}(S_n-1)$, is a problem in a ``replicated'' 
 classical 2D 
 critical systems  with a ``conformal defect'' along some curve,\cite{Oshikawa1997}  i.e a problem in boundary (Euclidean) conformal field theory (BCFT) on a 
 somewhat unusual manifold. The scaling behavior of $S_n$ with sub-system size is then not as surprising as it is strongly reminiscent of the finite-size scaling of the 
 free energy in large finite critical systems.\cite{Kac1966,Privman1984,CardyPeschel,Privman1988} On the other hand, it is known that the structure of the CFT 
 strongly determines its possible allowed BCFTs. Thus, the structure of the CFT that underlies these scale invariant wave functions must similarly play a key role in 
 the behavior of the entropies $S_n$. This natural relation of the entanglement entropy with boundary (or surface) critical behavior on a suitably defined manifold 
 appears naturally in the path integral formulation.\cite{Calabrese2004,Metlitski}

 In this work we use BCFT methods to reexamine this problem for the case of the quantum Lifshitz wave function studied before in Refs.\cite{Fradkin2006,Hsu2009} 
 paying close attention to the role of the compactification radius $R$ by constructing explicitly the boundary states of the associated BCFT.\cite{Cardy1989} Here we 
 show that in the limit of a large aspect ratio, $L \gg \ell \gg a$ (where $a$ is the short-distance cutoff and $L$ is the linear size of the full system) the entanglement 
 entropy has the universal finite term $\gamma_{QCP}=\ln R$, in agreement with the previous result of Ref.\cite{Hsu2009}.

 On the other hand, numerical estimates of the sub-leading corrections of the scaling of the entanglement entropy in the 2D QDM, obtained using the wave function 
 of the lattice model and using extrapolation methods, suggested that there may be additional universal contributions
 to $\gamma_{QCP}$.\cite{Furukawa2007,pasquier2009} These authors attribute the apparent disparity to the boundary conditions 
 used in Refs.\cite{Fradkin2006,Hsu2009}. This motivated us to reexamine the BCFT of this problem and to give an explicit form of the boundary states.
In BCFT, boundary 
 conditions are in one to one correspondence with the primaries of the bulk CFT.\cite{Cardy1989} 
For the case of multiple copies at hand here the bulk CFT possesses additional symmetries. The imposition of boundary conditions along a specified curve, a 
``conformal defect,''  reduces these symmetries.
A similar situation occurs when considering defect lines in the Ising model.\cite{Oshikawa1997}

A better understanding of systems of $n$-coupled critical systems is also crucial for reasons outside the entanglement entropy. From a broader viewpoint, a familiar 
condensed matter context are defects in lattice models. As mentioned earlier, one such example are defect lines in the Ising model.\cite{Oshikawa1997} More 
generally, conformal defects are very hard to classify. Even for two copies of a free bosonic theory, central charge $c=2$, 
the complete classification remains elusive.\cite{affleck2001} Theoretically, the classification of such defects is an important issue. These represent fixed points of a 
BCFT. A familiar example is the $k$-channel Kondo problem.\cite{Affleck2008} More complicated examples of $n$ intersecting theories also appear in a more 
applied context. The intersection of multiple quantum wires at the same point is one example.\cite{oshikawa2006} As mentioned earlier, systems of $n$-coupled 
degrees of freedom also make their appearance in the theory of quantum Brownian motion\cite{affleck2001,CaldeiraLeggett3} and in the dissipative Hofstadter 
model.\cite{hofstadter,callan1995,CallanFreed} In this work, we add the entanglement entropy in conformal quantum critical points to this list.

We compute the quantity $\textrm{Tr }\rho_{A}^{n}$ in terms of the original degrees of freedom and show that there is no additional factor as suggested 
by Ref.\cite{pasquier2009}. While our result agrees with our previous work, we note that this happens only in the asymptotic limit of a long cylinder and that for finite 
sized systems, there is generically a non-trivial $n$-dependence. To arrive at these conclusions, we identify a new boundary state describing the boundary condition 
that $n$ copies of a system are stitched together at an interface, and we argue that additional sub-leading corrections to Eq.(\ref{eq:SQCP}) are a result of differing 
boundary conditions used. We show that the correct boundary condition for the original degrees of freedom is one where the extra factor vanishes. This turns out to 
correspond to a subclass of conformal defects, at the common boundary. In addition, we show that there is a geometrical interpretation to
$\textrm{Tr }\rho_{A}^{n}$ as the ratio of classical partition functions defined on different tori, and that the universal sub-leading term is the ``thin torus limit'' of this 
ratio. With an explicit calculation, we find analytically that in the $n\rightarrow\infty$ the entanglement spectrum is given by the characters of the underlying conformal 
field theory describing the ground state wavefunction. This provides a case where it is possible to verify analytically the conjecture of 
Li and Haldane.\cite{Haldane2008} Finally, we demonstrate that in a string theory language, it is possible to think of the common boundary condition as a brane with 
a background gauge field and that $\textrm{Tr }\rho_{A}^{n}$ is the expectation value of an appropriately defined vertex operator (in the ``target space'' not in the CFT 
of the world-sheet).

The paper is organized as follows: in the next section, we review the construction of $\textrm{Tr }\rho_{A}^{n}$ for conformal quantum critical models. We then 
specialize to the quantum dimer model in Section \ref{sec:QDM} with the bulk of the calculation in Sec \ref{sec:calc}. A brief review of boundary conformal field theory 
in \ref{sec:bcft} is given as it plays a central role in the main calculation. Using the explicit expression for $\textrm{Tr }\rho_{A}^{n}$ we show in Section \ref{sec:es} 
that the entanglement spectrum has a level degeneracy indicative of the underlying conformal field theory describing the ground state wavefunction. In Section \ref{sec:Disc} we relate the boundary condition to a brane with a 
specific magnetic 
and electric fields and show that $\textrm{Tr }\rho_{A}^{n}$ can be computed as the expectation value of a vertex operator.

{\bf Note added to the text:} While this paper was being refereed, we received a preprint by Oshikawa \cite{oshikawa} and became aware of a number of inconsistencies in our treatment of the boundary state. In particular, it was pointed out that our glueing condition violated a permutation symmetry in the problem. These corrections and its consequences are noted in Section \ref{sec:note}.

\section{Conformal Quantum Critical Points}
\label{sec:rho_n}

We are interested in the von Neumann entropy for systems at conformal quantum critical points. These are systems where the norm of the ground state wavefunction is equivalent to a partition function of a two dimensional classical statistical model at criticality, \emph{i.e.} Eqn. (\ref{eqn:gs}). With the explicit form of the wavefunction, one can compute the entanglement entropy. This is the von Neumann entropy of the reduced density matrix
\begin{equation}
	S_{ent} = -\textrm{Tr } \rho_{A} \ln \rho_{A} = - \frac{\partial}{\partial n} \textrm{Tr} \rho_{A}^{n}.
\end{equation}
Since the ground state wave function is a local function of the field $\phi(x)$, a general matrix element of the reduced density matrix is a trace of the density matrix of the pure state $\Psi_{GS}[\phi]$ over the degrees of freedom of the ``unobserved'' region $B$, denoted by  $\phi^B(x)$. Hence the matrix elements of $\rho_A$ take the form
\begin{eqnarray}
\langle \phi^{A}_{i}  \vert  \hat{\rho}_{A}  \vert\phi^A_{i+1}   \rangle
= \frac{1}{Z} \int [D\phi^{B}_{i} ] \,\, e^{\displaystyle{-\left(\frac{1}{2} S^{A}(\phi^{A}_{i}) + 
\frac{1}{2} S^{A}(\phi^A_{i+1} ) 
+S^B(\phi^B_{i} )\right) } }  \, , \nonumber \\
\end{eqnarray}
where the degrees of freedom satisfy the {\em boundary condition} at the common boundary
 $\Gamma$:
\begin{equation}
\phi^B_{i} \vert_\Gamma=\phi^A_{i} \vert_\Gamma={{\phi}^A}_{i+1} \vert_\Gamma.
\label{eq:BCphiGamma}
\end{equation}
This problem can be thought of as a problem in boundary conformal field theory by letting $\lambda$ parameterize the boundary interaction between copy $i$ and $i+1$, at the strong coupling limit $\lambda\rightarrow \infty$ the copies are required to have the same configuration on the boundary and at $\lambda \rightarrow 0$ the copies do not interact at the boundary. Formally, the trace over $n$ copies of the reduced density matrix can be written as the ratio of the partition functions in these two limits,
\begin{equation}
	 \textrm{Tr} \rho_{A}^{n} =\frac{Z_{\lambda\rightarrow \infty}(n) }{Z_{\lambda\rightarrow 0}(n) } \label{eqn:rhoA}.
\end{equation}
The theoretical challenge is to compute this ratio. Various formal mathematical devices have been devised, but until now a direct approach has been lacking. It is desirable to understand the boundary condition in terms of the original degrees of freedom since those are most directly related to a physical dimer covering of the lattice.

\subsection{Quantum Dimer Model}
\label{sec:QDM}

Here we consider the simple case where the action in (\ref{eqn:gs}) is the Gaussian free field theory, described by a bosonic field with the property that it is identified on a circle of radius $R$, $\varphi\simeq\varphi+2\pi R$, 
\begin{equation}
	S = \frac{1}{8\pi} \int d^{2}x \,\, \partial \varphi \bar{\partial}\varphi \label{eqn:gs2}.
\end{equation}
One can think of this ground state (\ref{eqn:gs}) as a superposition lattice configurations in a statistical model such as the dimers on a square lattice \cite{Rokhsar1988} and its generalizations \cite{Castelnovo2005,Alet2005} or as coverings in the eight-vertex model \cite{Ardonne2004} with special choice of the Baxter weight \cite{Baxter1982}.

One can think of this action as describing a $1+1$ dimensional system defined on some manifold, a cylinder here. To connect with our pervious work, we let the circumference be of length $\ell$ and the length $L$. Periodic boundary conditions are imposed in the $\hat{x}$-direction. The field $\varphi(x,t)$ is a linear combination of the holomorphic $\phi(x,t)$ and anti-holomorphic $\bar{\phi}(x,t)$ parts, $\varphi(x,t) = \phi(x,t) +\bar{\phi}(x,t)$. Holomorphic and anti-holormophic mode expansions can be written as
\begin{eqnarray}
	\phi(x,t) &=& \varphi_{0} + \frac{2\pi }{\ell} \pi_{0} (t+ix) + \sum_{k\neq 0} \frac{i}{k} \alpha_{k} e^{\frac{2\pi  k}{\ell} (t+ix)} ,\nonumber \\
	&& \\
	\bar{\phi}(x,t) &= &\bar{\varphi}_{0} + \frac{2\pi }{\ell} \bar{\pi}_{0} (t-ix) + \sum_{k\neq 0} \frac{i}{k} \bar{\alpha}_{k} e^{\frac{2\pi  k}{\ell} (t-ix)},\nonumber \label{eqn:modeexpansion}
\end{eqnarray}
where $k\in \mathbb{Z}$ and the zero modes are given by
\begin{eqnarray}
	\pi_{0} = \left(\frac{m}{R} +\frac{wR}{2} \right) &;& \bar{\pi}_{0} = \left(\frac{m}{R} -\frac{wR}{2} \right) \label{eqn:zeromode}.
\end{eqnarray}
The constants $\varphi_{0}$ are canonically conjugate to these zero modes, $\left[ \pi_{0},\varphi_{0} \right]=i$.
The primaries of the boson field theory are labelled by the value of the zero modes, $(\pi_{0},\bar{\pi}_{0})$.

 \begin{figure} 
	 \includegraphics[width=0.4\textwidth]{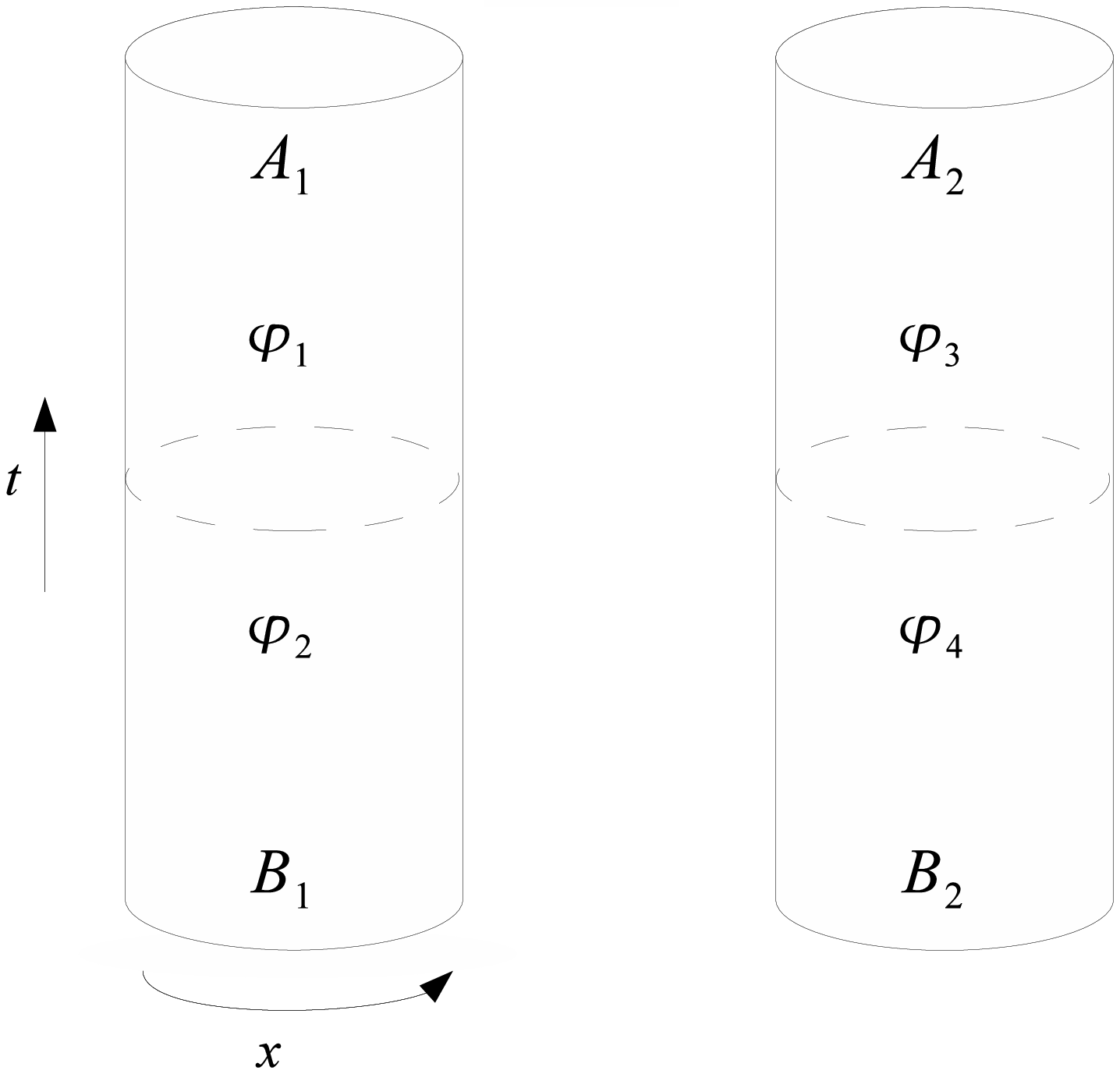}
\caption{We are interested in the long cylinder limit where $L \gg\ell$ where $L$ is the length of the cylinder and $\ell$ is the length of the circumference. \label{fig:copies}} 
\end{figure}

At the ends of the cylinder, Dirichlet boundary conditions can be chosen $\varphi(x,t=\pm L/2)=0$. On the oscillator modes, one finds that
\begin{equation} 
	\alpha_{k} = - \bar{\alpha}_{k}^{\dagger} q^{k} \label{eqn:dBC}.
\end{equation}
Here, $q=e^{2\pi i \tau}$ where $\tau = i \frac{L}{2\ell}$. For the zero modes, Dirichlet boundary conditions set the winding modes $w=0$. At the boundary $t=0$ continuity of the fields implies that $\lim_{a\rightarrow 0} \varphi(a) = \varphi(-a)$. This restricts the modes to obey
\begin{eqnarray}
	\alpha^{+}_{k} = \alpha^{-}_{k} \quad ; \quad \varphi_{0}^{+} = \varphi_{0}^{-} \quad ; \quad \pi_{0}^{+} = \pi_{0}^{-} \label{eqn:continuity}, \label{eqn:cBC}
\end{eqnarray}
which amounts to simple continuity of the field at the boundary in agreement with (\ref{eq:BCphiGamma}). Now, we choose to observe the lower half of the cylinder ($t\leq0$), region $B$, and compute the reduced density matrix for the remaining region $A$ by the replica trick. The fields at the boundary should be continuous with each other as described in (\ref{eq:BCphiGamma}) and follow an explicit relationship among the modes given by (\ref{eqn:continuity}). Computing the reduced density matrix is hence a computation of a ratio of partition functions (\ref{eqn:rhoA}). This is most simply done in the boundary conformal field theory framework. This formalism is briefly reviewed in \ref{sec:bcft}. As it will play a central role, those unacquainted should proceed there. More extensive reviews of the subject matter can be found in other canonical texts.\cite{bigyellowbk}

\subsection{Boundary States}
\label{sec:calc}

To make use of boundary conformal field theory, we fold the system at the boundary at $t=0$ (see Figure \ref{fig:copies}) so that there are $2n$ cylinders of half the total length $L$. At $t=L/2$, Dirichlet boundary conditions were imposed on the ends which relates the holomorphic and anti-holomorphic modes. The boundary condition (\ref{eqn:dBC}) can be regarded as an eigenvalue expression for the annihilation operators, and the Ishibashi states can be written as the coherent state
\begin{eqnarray}	
	\vert m \rangle \rangle =  \prod_{k=1} \exp\left( \bar{\alpha}^{i^{\dagger}}_{k} Q_{ij} \alpha^{j^{\dagger}}_{k} \right) \vert m;0 \rangle,
\end{eqnarray}
where and $\alpha^{i}_{k} \vert m;0 \rangle=0$ and the state $\vert m;0\rangle$ labels states in the Fock space where the winding mode $w=0$, as required by Dirichlet boundary conditions (\ref{eqn:dBC}). With two copies, the matrix $Q_{ij}$ is given by, $Q_{ij} = -q^{k}\delta_{ij}$.
The boundary state is a linear combination of these Ishibashi states, given by (\ref{eqn:bstate}) and (\ref{eqn:bosonBstates})
\begin{equation}
	\vert B_{D} \rangle = g_{D}  \sum_{m=-\infty}^{\infty} e^{i \frac{\hat{m}}{R} \varphi_{0} } \vert m \rangle \rangle \label{eqn:DBstate}.
\end{equation}
$g_{D}$ is the $g$-factor associated with the Dirichlet boundary condition for the free boson and can be computed explicitly as mentioned in (\ref{eqn:gfactor}).

We now construct the non-trivial boundary state that is at the boundary $t=0$. Below, we specialize to the case of $n=2$ copies, but the result is easily generalized to arbitrary $n$. At $\lambda \rightarrow 0$, the two copies are decoupled, and the boundary condition becomes
\begin{eqnarray}
	\alpha^{1^{\dagger}}_{k} = \bar{\alpha}^{2}_{k} \quad&;&\quad \alpha^{2^{\dagger}}_{k} = \bar{\alpha}^{1}_{k} \nonumber \\
	\alpha^{3^{\dagger}}_{k} =\bar{\alpha}^{4}_{k} \quad&;&\quad \alpha^{4^{\dagger}}_{k} = \bar{\alpha}^{3}_{k}. \label{eqn:uncoupled}
\end{eqnarray}
and the momentum modes are restricted to obey,
\begin{eqnarray}
	\pi_{0}^{1}=\bar{\pi}_{0}^{2} \quad&;&\quad \pi_{0}^{3}=\bar{\pi}_{0}^{4} \nonumber \\
	\bar{\pi}_{0}^{1}=\pi_{0}^{2} \quad&;&\quad \bar{\pi}_{0}^{3}=\pi_{0}^{4}. \label{eqn:pfree}
\end{eqnarray}
The equation (\ref{eqn:uncoupled}) can be regarded as eigenvalue equations for the destruction operators and the Ishibashi state for each can be written succinctly as
\begin{eqnarray}
	\,_{\lambda}\langle\langle m';w' \vert = \langle m';w'\vert \prod_{k=1} \exp \left( \bar{\alpha}^{i }_{k}\mathcal{R}_{ij}(\lambda) \alpha^{j}_{k} \right), 
\end{eqnarray}
where
\begin{equation}
	\mathcal{R}_{ij}(\lambda\rightarrow0) = \left( \begin{array}{cccc} 0&1&0&0 \\ 1&0&0&0 \\ 0&0&0&1 \\ 0&0&1&0 \end{array}\right).\label{eqn:Pfree}
\end{equation}
A similar construction can be done for the $\lambda\to\infty$ case. In a previous version of the paper, $\mathcal{R}_{ij}(\lambda\to\infty)$ was given incorrectly and was pointed out by Oshikawa \cite{oshikawa} to violate a permutation symmetry of the problem. Instead all possible boundary conditions must be summed (see Section \ref{sec:note}). The resulting matrix is given by equation (\ref{eqn:matrix}).

The state $\langle m';w'\vert$ denote states with eigenvalues obeying the matrix relationship 
\begin{equation}	
	\pi_{i} = \mathcal{R}_{ij} \bar{\pi}_{j}
\end{equation}
where the form of $\mathcal{R}_{ij}$ depends on the value of $\lambda$. It is given by (\ref{eqn:Pfree}) for $\lambda\to0$ and (\ref{eqn:matrix}) for $\lambda\to\infty$. Hence, the boundary state can be written as a linear combination of Ishibashi states again,
\begin{equation}
	\langle B_{\lambda} \vert = \,_{\lambda}\langle\langle m';w' \vert \,\, g_{\lambda} \sum e^{i\hat{\pi'}_{0} \varphi_{0} }.
\end{equation}
where the sum is over modes on the correct compactification lattice. This subtlety was neglected in a previous version of the paper, but was pointed out by Oshikawa \cite{oshikawa}. Here, the $g$-factor will play an important role. We call the $g$-factor associated with the $\lambda\rightarrow\infty$ limit $g_{UV}$ and the $g$-factor associated with the $\lambda\rightarrow 0$ limit $g_{IR}$. We leave the specific evaluation of these normalization factors till later.

The partition function is then easily evaluated as the overlap of the two boundary states. 
\begin{eqnarray}
	Z_{\lambda}(n) &=& \langle B_{\lambda} \vert q^{\hat{H} } \vert B_{D} \rangle  =  g_{\lambda} g_{D} q^{-2n/12}\langle \varphi'_{0} \vert q^{\hat{H}(\lambda)}\vert \varphi_{0} \rangle Z_{osc}, \nonumber \\
	&&
\end{eqnarray}
where the quantity $\langle \varphi'_{0} \vert q^{\hat{H}}\vert \varphi_{0}\rangle $ is the piece only involving the zero modes and $Z_{osc}$ involves the oscillator modes. The respective matrices describing the glueing conditions inserted, one readily finds,
\begin{eqnarray}
	 Z_{osc}(\lambda\to\infty) = \prod_{k>0}  \left( 1 - (q^{k} \right)^{-(2n-1)} (1+q^{k} )^{-1} , \nonumber \\
	Z_{osc}(\lambda\to0) = \prod_{k>0} \left( 1 - (q^{2})^{k} \right)^{-n}.
\end{eqnarray}


 \begin{figure*}
	\includegraphics[width=\textwidth]{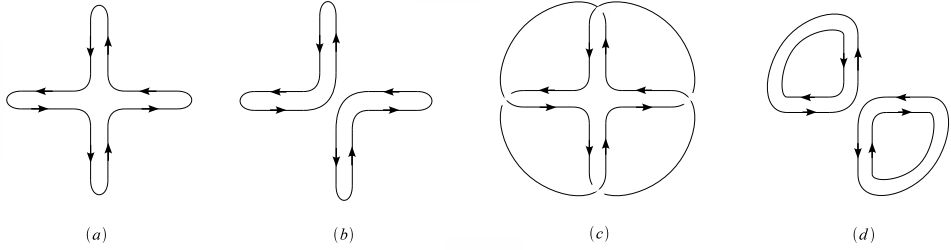}
	\caption{In the two limits $\lambda\rightarrow \infty$ (a),(c) and $\lambda\rightarrow0$ (b),(d), the holomorphic (ingoing arrows) and anti-holomorphic (outgoing arrows) are stitched in topologically distinct ways. The legs are labelled $\varphi_{1},\varphi_{2},$ etc. in a counterclockwise way. In (a), there is effectively only one bosonic degree of freedom, while in (b) they are stitched for form $n$-independent closed loops associated with an independent bosonic degree of freedom. In figure (c),(d) a similar diagram can be drawn to compute the $g$-factors. In (c), $\lambda\rightarrow\infty$, there is only one non-chiral bosonic degree of freedom while in (d), $\lambda\rightarrow 0$, there are two non-chiral bosonic degrees of freedom. Because of the permutation symmetry, one must sum over all possible closed, chiral loop configurations. This corresponds to a sum over glueing conditions in Section \ref{sec:note} and the glueing matrix given by Oshikawa \cite{oshikawa}.\label{fig:glue}}
\end{figure*}

The zero modes require some care. In a previous version of this paper, the winding and momentum modes were assumed to be independent, but the glueing matrix forces a enforces a condition. Instead one must sum over a lattice $\Xi$ as noted by Oshikawa \cite{oshikawa}. This is dual lattice which satisfies the glueing condition on the zero modes,
\begin{equation}
	\pi_{i} = \mathcal{R}_{ij} \bar{\pi}_{j}
\end{equation}
where $\mathcal{R}_{ij}$ is given by (\ref{eqn:matrix}). Hence for $\lambda\rightarrow\infty$, one finds,
\begin{equation}
	\langle \varphi'_{0} \vert q^{\hat{H}(\lambda\rightarrow\infty)} \vert \varphi_{0}\rangle= \sum_{\vec{K} \in \Xi} \tilde{q}^{\vec{K}^{2} /2} 
\end{equation} 
By a similar set of arguments, for $\lambda\to 0$ one finds,
\begin{equation}
	\langle \varphi'_{0} \vert q^{\hat{H}(\lambda\rightarrow0)} \vert \varphi_{0}\rangle = \left(\sum_{m=-\infty}^{\infty} (q^{2} )^{\frac{1}{2} \frac{m^{2}}{R^{2}}  } \right)^{n}.\label{eqn:zmodes_0}
\end{equation}
The results are easily generalized to arbitrary $n$. Writing in terms of $\vartheta_{3}$-functions (see \ref{sec:AppendixTheta}), the UV and IR limit partition functions are given by,
\begin{eqnarray}
	Z_{\lambda\rightarrow\infty} (n) &=& g_{UV}g_{D} \left( \frac{1}{\eta(\tilde{q} )} \right)^{2n-1} \tilde{q}^{-1/24} \prod_{m=1}^{\infty} \frac{1}{1+\tilde{q}^{m} }  \sum_{\vec{K} \in \Xi} \tilde{q}^{\vec{K}^{2} /2} \label{eqn:Zinfty}\\
	Z_{\lambda\rightarrow 0}(n) &=& g_{IR} g_{D}\left(\frac{   \vartheta_{3}\left(0\vert \frac{2\tau}{R^{2} }\right)}{ \eta(2\tau)} \right)^{n}  .\label{eqn:Zzero} 
\end{eqnarray}
As a non-trivial check, for $Z_{\lambda\rightarrow 0}$ we note that $2\tau$ is the modular parameter for a cylinder of total length $\ell$ and observe that the term in paranthesis is the $U(1)$ character; it is the partition function for $n$-decoupled bosons on a cylinder of length $\ell$ as it should be. In light of this observation, $g_{IR}$ should be equal to 1.

The factors $g_{UV}$ and $g_{IR}$ are normalization factors that have yet to be fixed. These can be fixed by Cardy's conditions and one finds that, 
\begin{equation}
	g_{IR} = 1, \quad 	g_{UV}=R^{(1-n)}\sqrt{n}
\end{equation}

Crucially, the ratio $g_{UV}/g_{IR}=R^{(1-n)}\sqrt{n}$, and hence we arrive at the main result of this paper,
\begin{eqnarray}
	\textrm{Tr }\rho_{A}^{n} = C(n,\ell,L)  \frac{g_{UV} }{g_{IR}}\left( \frac{Z_{\lambda\to\infty} }{Z_{\lambda\to0}} \right) \label{eqn:result1}
\end{eqnarray}
where $Z_{\lambda\to\infty}$ and $Z_{\lambda\to 0}$ are given by (\ref{eqn:Zinfty}) and (\ref{eqn:Zzero}) respectively and we have factored out the constant factors $g_{UV}/g_{IR}$ explicitly. Here, we have included  the regulator dependent contribution in the function $C(n,\ell,L)$ that comes from the short distance cutoff which we have hitherto suppressed.\cite{Weisberger87} We comment on its specific form below. More importantly, we note that the each of terms $\vartheta_{3}/\eta$ are characters of a $U(1)$, $c=1$ conformal field theory.\cite{bigyellowbk}

Thus far, we have been concerned only with the regulated part of $\textrm{Tr }\rho_{A}^{n}$ and have neglected the divergent contributions from the short distance cutoff. These can be recovered by a careful $\zeta$-function regularization.\cite{Weisberger87} In general, it was shown that in the limit of interest the free energy for a system on a \emph{smooth} open manifold scales as,\cite{CardyPeschel}
\begin{equation}
	\ln Z_{\lambda} = \mu_{a} \vert A \vert + \mu_{b} P -\frac{c}{6} \chi \ln \ell + \mathcal{O}(1) +\mathcal{O}(\ell/L). 
\end{equation}
Here $A$ is the area of the cylinder, $P$ the perimeter of the boundary and $\chi$ is the Euler character of the manifold (zero for cylinders). The coefficients $\mu_{a}$ and $\mu_{b}$ depend on the short distance cutoff. The order one contributions are what have been computed explicitly thus far. We note in both limits, the total area is simply $n A$ where $A$ is the area of a single cylinder so that in the ratio (\ref{eqn:rhoA}) this contribution cancels. The difference in the two limits lies in the perimeter of the boundary. In the decoupled limit, there are $2n$ unshared boundaries of length $\ell$ and $n$ shared (where the two halves are joined) boundaries so that the total perimeter of the boundary is $(n+2n)\ell$. Meanwhile, in the strongly coupled regime all the manifolds  coincide (smoothly) on a single interface of length $\ell$ and there are still the same $2n$ unshared boundaries. The total length of the perimeter in this case is $(1+2n)\ell$. The important point is that there is also a divergent non-universal contribution to $\textrm{Tr }\rho_{A}^{n}$. To leading order, 
\begin{equation} 
	 C(n,\ell,L) = e^{\mu_{b} (1-n)\ell} .
\end{equation}
where the prefactor $\mu_{b}$ is non-universal  and  depends on the short distance regulator.\cite{Weisberger87} It is important to note that this analysis neglects the effects cusps and corners that occur if the boundary is not smooth, giving rise to conical singularities, but these have been shown to give $\ell$ dependent contributions that scale as a power law that may be non-trivial functions of $n$ and to additional universal (logarithmic) corrections to the entanglement entropy.\cite{Fradkin2006,CardyCalabrese2010}

It is important to observe that as $n\rightarrow1$, $\textrm{Tr }\rho_{A}=1$ as it should be. Now, in the present long cylinder limit the $\vartheta_{3}$-functions and $\eta$-function are equal to the identity to leading order in $e^{-2\pi i /\tau}$. Our main result is that the universal sub-leading term to the entanglement entropy is,
\begin{equation}
	\gamma_{QCP} = \ln R-\frac{1}{2}, \label{eqn:result3}
\end{equation}
where, once again, $R$ is the compactication radius of the boson. This agrees with numerical result for the quantum dimer model.\cite{pasquier2009}

\section{Discussion}
While the result for $\gamma_{QCP}$ is the same as our previous result,\cite{Hsu2009} the interpretation is much different. The full analytic expression for $\textrm{Tr }\rho_{A}^{n}$, for instance, differs and it is only in the asymptotic limit that the two results are identical. In the very long cylinder limit, $L/\ell \rightarrow\infty$ the characters asymptotically approach the identity, and one finds that $\textrm{Tr }\rho_{A}^{n}$ identically reduces to the one obtained previously.\cite{Hsu2009} However, for large but finite values of the aspect ratio $L/\ell$,  $\textrm{Tr }\rho_{A}^{n}$  has a non-trivial $n$-dependence. Naturally, an immediate question is what information is contained in $\textrm{Tr }\rho_{A}^{n}$ in this picture? We show below that the entanglement spectrum is indeed described by the underlying conformal field theory describing the ground state wavefunction. More importantly, several immediate questions arise. First, the form of $\textrm{Tr }\rho_{A}^{n}$ is relatively simple and one wonders if there is a deeper reason for this. Secondly, we have constructed boundary states $\vert B_{\lambda\rightarrow 0}\rangle$ and $\vert B_{\lambda\rightarrow \infty}\rangle$ which lack a straightforward classification as free or fixed. They represent instead a coupling between different copies. An understanding of these states is clearly desirable. Lastly,  our result for $\gamma_{QCP}$ hinged on $g_{UV}/g_{IR}=R^{(1-n)}\sqrt{n}$ and a natural question is whether other values are possible. In what follows, we address each of these issues.

\subsection{The Entanglement Spectrum}
\label{sec:es}
With an explicit expression for $\textrm{Tr }\rho_{A}^{n}$, one in fact can construct all the moments of the R\'enyi entropy,
\begin{equation}
	S_{n} = \frac{1}{1-n} \ln \, \textrm{Tr }\rho_{A}^{n} .
	\label{eq:Sn}
\end{equation}
One can examine the degeneracy of states by examining the higher moments, $n\rightarrow\infty$, of the R\'enyi entropy. It has been postulated, but not shown analytically, that the higher moments of the R\'enyi entropy should be given by the characters of the underlying conformal field theory describing the ground state wavefunction.\cite{Haldane2008} The characters of a conformal field theory count the number of independent states occurring at a given energy level. Here, we find that in the $n\rightarrow\infty$ limit the contribution from the strong $\lambda\rightarrow\infty$ coupled sector tends to the identity so that the R\'enyi entropy in this limit is given by,
\begin{equation}
	S_{n\to\infty} = \frac{1}{1-n} \log \, \textrm{Tr }\rho_{A}^{n} = \frac{1}{1-n} \left( -\frac{1}{12} \frac{\pi L}{\ell} + \log \phi(\tilde{q}) \right) + \dots
	\label{eq:Sn-limit}
\end{equation}
where the $\dots$ indicate subleading constant contributions. Here, $\phi(\tilde{q})$ is the Euler function that is related to the $\eta$-function through Ramanujan's identity. 
\begin{equation}
	\phi(\tilde{q}) = \tilde{q}^{-c/24} \eta(\tilde{q}).
\end{equation}
$1/\phi(\tilde{q})$ is related to the partitions of integers, $p(k)$, and $\tilde{q} = e^{-2\pi i /\tau}$ where $\eta(-1/2\tau) = \eta(\tilde{q})$,
\begin{equation}
	\frac{1}{\phi(\tilde{q}) } = \sum_{k=0}^{\infty} p(k) \tilde{q}^{k}
\end{equation}
In the long cylinder limit, $\tilde{q}$ is exponentially small and $\theta_{3}\rightarrow 1$. One finds that,
\begin{equation}
	S_{n\rightarrow\infty}(L\gg\ell) = \mu_{b}\ell  - \frac{\pi c}{12} \frac{L}{\ell} - \ln \left(\sum_{k=0}^{\infty} p(k) \tilde{q}^{k}\right) .
\end{equation}
Asymptotically the multiplicities are given by the the partition of integers which exactly describes the number of states at a given energy in a $c=1$ free bosonic conformal field theory. Therefore, in a finite size computation of the entanglement spectrum the degeneracies of the eigenvalues of the reduced density matrix are (asymptotically) given by the integers $p(k)$. This relation was conjectured (for quantum Hall wave functions) by Li and Haldane.\cite{Haldane2008}

\subsection{Background Electromagnetic Fields}
\label{sec:Disc}
One issue still untouched by the preceding discussion is where additional universal corrections to $\gamma_{QCP}$ might come from. Some insight can be gained by realizing that the problem of $2n$ free bosonic field theories interacting only at a common boundary has been studied in the context of quantum Brownian motion in a magnetic field,\cite{affleck2001} and open strings in a background gauge field. \cite{callan1995,Callan1986} The connection can be seen more concretely by considering $\textrm{Tr }\rho_{A}^{n}$ in the path integral formulation
\begin{equation}
	\textrm{Tr }\rho_{A}^{n} = \frac{ \int D\varphi \, e^{-S_\infty[\varphi] } }{\int D\varphi \, e^{-S_0[\varphi]} },
	\label{eq:newrep}
\end{equation}
where $S_\infty[\varphi]=\lim_{\lambda \to \infty} S_\lambda[\varphi]$ and $S_0[\varphi]=\lim_{\lambda \to 0}S_\lambda[\varphi]$ describes the bosonic action of $n$ scalar fields  satisfying boundary conditions specified by $\lambda$. Once again, the numerator describes $n$ fields that are forced to coincide at the boundary (hence $\lambda \to \infty$) and in the denominator the $n$ fields are decoupled from each other (hence $\lambda \to 0$).

We will now write $S[\varphi]$ in a form that we find more useful as follows. The first step in understanding this problem in terms of quantum Brownian motion in a magnetic field or, equivalently, open strings in a background gauge field, is to fold the system accross the boundary, thus doubling the number of fields. Let $\Phi_i$, with $i=1,\ldots,2n$ denote a $2n$ component scalar field whose upper $n$ components label the (folded) fields from the $A$ regions and its remaining (lower) $n$ components are those of the $B$ region,
\begin{equation}
\Phi^i=\left( \varphi^1_A,\ldots,\varphi^n_A,\varphi^1_B,\ldots,\varphi^n_B\right) .
\end{equation}
The action for the $\Phi$ field is (for so far unspecified boundary conditions)
\begin{equation}
	S[\Phi] = \frac{1}{2} \int d^{2}x \, \sum_{i=1}^{2n} \left(\partial_{\mu}\Phi^{i} \right)^{2} .
	 \label{eqn:gauged}
\end{equation}
Here we choose our coordinates so that $x$ is the direction parallel to the length of the cylinders and $t$ runs along the circumference.

Now we make a ``$T$-dual" transformation on the $B$ field, $\tilde{\varphi_B}^{1}, \,.\,.\,. \tilde{\varphi_B}^{n}$. This corresponds to a symmetry of the action (on the $B$ fields) with respect to the interchange of their winding and charge modes,\cite{bigyellowbk,gswv1}
\begin{equation}
	\begin{array}{ll}
		m^{i} \rightarrow w^{i}, & w^{i}\rightarrow m^{i} \\
		\alpha^{i}_{n} \rightarrow \alpha^{i}_{n}, & \bar{\alpha}^{i}_{n} \rightarrow - \bar{\alpha}^{i}_{n},
	\end{array}
\end{equation}
Here $m^i$ and $w^i$ label the winding and charge numbers of the zero modes of the fields $\varphi_B^i$. Under $T$-duality the compactification radius $R$ transforms as $R\rightarrow 2/R$, while Neumann boundary conditions transform into Dirichlet boundary conditions, and viceversa.

The fields only interact with each other at the common boundary via the boundary conditions. To this end we introduce a field $A_{i} = \frac{1}{2} F_{ij} \Phi^{j}$ localized at the boundary,
\begin{equation}
	S[\Phi^{i} ] = \frac{1}{2}\int d^{2}x \, \,  \sum_{i=1}^{2n} \left( \partial_{\mu} \Phi^{i} \right)^{2} -  \oint dt \, A_{i} \partial_{t} \Phi^{i}.
\end{equation}
where $F_{ij}$ is an antisymmetric matrix we define below.
Formally, this action describes the ``dissipative Hofstadter model'' \cite{CallanFreed} (with vanishing potential).  Upon varying the action, the fields $\Phi^{i}$ are found to obey the usual wave equation with the boundary condition,
\begin{equation}
	\partial_{x} \Phi^{i} = F_{ij} \partial_{t} \Phi^{j}. \label{eqn:bcF}
\end{equation}
If $F_{ij}$ is a constant matrix, independent of $\Phi^{i}$, and anti-symmetric, then clearly $F_{ij} = \partial_{i}A_{j}-\partial_{j}A_{i}$. One can think of $F_{ij}$ as the 2-form field strength tensor associated with a gauge field $A_{i}$. Now, letting $F_{ij}$ be the $2n\times 2n$ matrix,
\begin{equation}
	F_{ij} = \left( 
	\begin{array}{cc}
		0 & -M^{T}_{mn} \\
		M_{mn} & 0
	\end{array}
	\right) . \label{eqn:F}
\end{equation}
it is readily seen that Eq.(\ref{eqn:bcF}) yields the desired boundary conditions at the common boundary for the scalar fields by a suitable choice of the $n \times n$ matrix $M$.

This construction can be used to represent both the numerator and denominator of Eq.\eref{eq:newrep} by the choices (in this basis)
\begin{equation}
M_\infty=
	\left(
\begin{array}{ccccc}
1 & 2 & 2 & \ldots & 2\\
2 & 1 & 2 & \ldots & 2\\
2 & 2 & 1 & \ldots & 2\\
. &  .  & .  & \ldots & .\\
. &  .   &  . & \ldots & 2\\
2 & .   & .  & \ldots& 1
\end{array}
\right),
\quad
M_0=
\left(
\begin{array}{ccccc}
1 & 0 & 0 & \ldots & 0\\
0 & 1 & 0 & \ldots & 0\\
0 & 0 & 1 & \ldots & 0\\
. &  .  & .  & \ldots & .\\
. &  .   &  . & \ldots & 0\\
0 & .   & .  & \ldots& 1
\end{array}
\right) ,
\label{eq:matrices}
\end{equation}
for $\lambda \to \infty$ and $\lambda \to 0$ respectively. Previously, a matrix was given for $M_{\infty}$ that violated the permutation symmetry present in the boundary condition. Summing over possible glueing conditions yields the matrix above. The matrix $M_{\infty}$ is then identical to the matrix $\mathcal{R}$ (see equation (\ref{eqn:matrix})) of Oshikawa \cite{oshikawa} up to an overall unimportant sign.

In analogy with electromagnetism, there is a electric and magnetic field at the boundary $x=0$ that mediates the interaction between the $2n$-dimensional systems. In terms of quantum Brownian motion,\cite{affleck2001} one can regard the problem as a particle at $x=0$ on a $2n$-dimensional space moving in a electric and magnetic field. As an open string, one sees that the boundary can be thought of as a brane carrying a magnetic and electric field.\cite{Callan1986,callan1995} The analogy also gives a possible interpretation of the result obtained by Ref. \cite{pasquier2009}. In the context of quantum Brownian motion in $n$-dimensions, it has been shown that there is a plethora of possible boundary states where $g_{UV}/g_{IR}$ is non-trivial. These boundary states correspond to a different electromagnetic field and hence a different coupling of the replicas in our picture.

We have thus constructed a gauge field (and associated 2-form) that describe the boundary interaction. One can go further by rewriting $\textrm{Tr }\rho_{A}^{n}$ in the suggestive form
\begin{equation}
	\textrm{Tr }\rho_{A}^{n} = \frac{\left\langle e^{\oint d\Phi^{i}  A_{i}^\infty  } \right\rangle }{  \left\langle e^{\oint d\Phi^{i} A_{i}^0   } \right\rangle} .
	\label{eqn:wilson}
\end{equation}
where $A_i^\infty$ corresponds to the choice $M_\infty$ and $A_i^0$ to $M_0$ (both defined above). The universal sub-leading correction to the entanglement entropy is the asymptotic behavior of this correlation function. In string theory, such objects are generically called vertex operators.\cite{gswv1} This operator can be understood as counting a topological charge
\begin{equation}
	Q =\frac{1}{2} \int d^{2}x \, F_{ij} \epsilon_{\mu\nu} \partial^{\mu}\Phi^{i} \partial^{\nu}\Phi^{j} .
\end{equation}
where the field $\Phi^{i}$ is a map, $\Phi^{i}: T^{2}\rightarrow T^{2n}$. Integrating by parts yields the correct boundary field. Importantly, there exists a basis of the field $\Phi^i$ where $F_{ij}$ is an $n$-block diagonal matrix of anti-symmetric $\epsilon_{ij}$ tensors. $\Phi^{i}$ can then be written as a tensor product of maps $T^{2}\rightarrow T^{2}$ and $Q$ can be thought of as a product of holonomies characterized by the homotopy group $\pi_{T^{2}}(T^{2})=\mathbf{Z}$. An interesting observation is that the gauge field $A_{i}$ is fixed by the free part of the action of Eq.(\ref{eqn:gauged}) so that a perturbation that brings the system into a topological phase, e.g.  $m \cos(\varphi),$ has no effect on the boundary condition and hence $A_{i}$ remains the same. We expect that this correlation function, and hence the entanglement entropy, should take different values in the different phases.

\section{Note Added to the Text}
\label{sec:note}
While this paper was being put into production, we received a preprint by Oshikawa \cite{oshikawa} and we became aware of a number of inconsistencies in our previous work. In particular, the matrix used to described the coupled boundary condition did not possess the full permutation symmetry of the problem and the winding and momentum quantum numbers of the zero modes were not treated correctly. 
As noted by Oshikawa, \cite{oshikawa} there is a permutation symmetry in the problem and each copy and definition of region A and B are interchangeable. One should then sum over all possible combinations of matrices $P_{ij}(\lambda\rightarrow\infty)$, where $P_{ij}(\lambda\rightarrow\infty)$ is a shift matrix. The glueing condition is then,
\begin{equation}
	\mathcal{R}_{ij} = \mathbf{1}+ ( P_{ij} + P_{ij}^{\dagger} ) + P_{ij}^{2} + P^{2\,\, \dagger}_{ij} + \dots + P^{(2n-1)}_{ij} +P^{(2n-1)\,\, \dagger}_{ij} 
\end{equation}
The resulting matrix $\mathcal{R}$ is in fact the matrix $\mathcal{R}$ given by Oshikawa \cite{oshikawa} up to an overall unimportant sign. Explicitly,
\begin{equation}
	\mathcal{R}_{ij} =
	\left(
\begin{array}{ccccc}
1 & 2 & 2 & \ldots & 2\\
2 & 1 & 2 & \ldots & 2\\
2 & 2 & 1 & \ldots & 2\\
. &  .  & .  & \ldots & .\\
. &  .   &  . & \ldots & 2\\
2 & .   & .  & \ldots& 1
\end{array}
\right) \label{eqn:matrix}
\end{equation}

Using the sum of all possible glueing conditions because of the permutation symmetry, the partition function (\ref{eqn:Zinfty})  is given by the expression found by Oshikawa \cite{oshikawa}.
\begin{equation}
	Z_{\lambda\rightarrow\infty}= g_{UV}g_{D} \left( \frac{1}{\eta(\tilde{q} )} \right)^{2n-1} \tilde{q}^{-1/24} \prod_{m=1}^{\infty} \frac{1}{1+\tilde{q}^{m} }  \sum_{\vec{K} \in \Xi} \tilde{q}^{\vec{K}^{2} /2} 
\end{equation}
Here, $\Xi$ is a lattice that satisfies the boundary condition on the zero modes,
\begin{equation}
	\pi_{i} = \mathcal{R}_{ij} \bar{\pi}_{j}
\end{equation}
The result for $Z_{\lambda\rightarrow 0}$ is unchanged and given by equation (\ref{eqn:Zzero}).  As pointed out by Oshikawa \cite{oshikawa}, the correction normalization is,
\begin{equation}
	g_{UV} = R^{1-n}\sqrt{n}
\end{equation}
The universal subleading term to the entanglement entropy is dependent on this ratio of $g$-factors so that the result (\ref{eqn:result3}) is now
\begin{equation}
	S = \log R -\frac{1}{2}
\end{equation}

Several results in the paper are modified because of this. In particular, the quantity $\textrm{Tr }\rho_{A}^{n}$ (i.e. equation (\ref{eqn:result1}) ) is now given by
\begin{equation}
	\textrm{Tr }\rho_{A}^{n} =  \frac{ g_{D} g_{UV}  \left( \frac{1}{\eta(\tilde{q} )} \right)^{2n-1} \tilde{q}^{-1/24} \prod_{m=1}^{\infty} \frac{1}{1+\tilde{q}^{m} } \sum_{\vec{K} \in \Xi} \tilde{q}^{\vec{K}^{2} /2} }{g_{D} g_{IR}\left( \frac{1}{\eta(\tilde{q})}\right)^{2n} \sum_{n=-\infty}^{\infty} \tilde{q}^{n^{2} R^{2}/4} } \label{eqn:corr}
\end{equation}
Then in the limit $L\gg\ell, \tilde{q}\rightarrow 0$ one finds that the distribution of eigenvalues of the R\'enyi entropy for large moments $n$, is given by
\begin{equation}
	S_{n\to\infty} = \frac{1}{1-n} \log \, \textrm{Tr }\rho_{A}^{n} = \frac{1}{1-n} \left( -\frac{1}{12} \frac{\pi L}{\ell} + \log \phi(\tilde{q}) \right) + \dots
\end{equation}
where $\phi(\tilde{q})$ is the partition of integers and $\dots$ include subleading constant terms. Eigenvalues are distributed according the level counting for the free $c=1$ boson theory that describes the ground state wavefunction, as mentioned in the body of the paper.

\section{Conclusion}
Before concluding with broader and more speculative issues, we summarize our results for the quantum dimer model. We found that in the limit $L\gg \ell$, $\textrm{Tr }\rho_{A}^{n}$ is given by the expression of Eq.(\ref{eqn:corr}). This was done in terms of the original degrees of freedom and it is hoped that this clarifies what boundary condition must be used at the common interface. In the limit $n\rightarrow 1$ of $S=-\partial_{n} \textrm{Tr }\rho_{A}^{n}$, it was found that the universal finite part of the entanglement entropy is
\begin{equation}
	\gamma_{QCP} = \ln R-\frac{1}{2}.
\end{equation}
This result coincides with the numerical result. \cite{pasquier2009} We found that for finite sized systems, there is a non-trivial $n$-dependence, not reflected in the construction of Fradkin and Moore.\cite{Fradkin2006} The source of this difference laid in the subtle details of defect lines in critical systems. In this work, we showed that the boundary condition is not described by the same notion of Dirichlet boundary conditions as in the original system. We circumvented the difficulties here by working directly with the original degrees of freedom.

By considering the original degrees of freedom, we further demonstrated that in the limit $n\rightarrow\infty$, $\textrm{Tr }\rho_{A}^{n}$ has a distribution of eigenvalues characterized by the correct underlying conformal field theory of the ground state wavefunction, confirming, at least for this case, a conjecture put forth by Li and Haldane.\cite{Haldane2008} Attempting to understand where the universal corrections to entanglement entropy come from, we related the problem formally to work done on quantum Brownian motion and branes with a background electromagnetic field, and showed that  $\textrm{Tr }\rho_{A}^{n}$ can be understood as an expectation value of a vertex operator.

In this work we focused on the quantum dimer model, \emph{i.e.} models where the norm of the ground state wavefunction is related to the free Gaussian field theory, but the methods can be readily extended to different conformal quantum critical models. It would be interesting to see if a similar structure exists for more complicated systems, such as non trivial topological theories with non-Abelian excitations.\cite{Freedman2004,Levin2005,Fendley2006,Fendley2008} For some simple cases, an exact solution is possible.\cite{Hsu2010b}

\ack
We thank Paul Fendley, Duncan Haldane, Vincent Pasquier, and Jean-Mar{\'\i}e St{\'e}phan for discussions. We especially thank Masaki Oshikawa for sending us his pre-print and discussing his results with us. This work was supported in part by the National Science Foundation grant DMR 0758462 at the University of Illinois.

\appendix

\section{Boundary Conformal Field Theory}
 \label{sec:bcft}

To exploit conformal invariance, it is useful to think of the system as being on a cylinder with circumference $\beta$ and length $\ell$ with boundary conditions $A,B$ on the field on the left and right ends of the system respectively. Quantum mechanically, this corresponds to evolving a one dimensional system defined on the line $x$ in time $\beta$. The partition function is given by the usual expression,
\begin{equation}
	Z_{AB} = \textrm{Tr } e^{-\beta H^{\ell}_{AB} }. \label{eqn:ostring}
\end{equation}

If the Hamiltonian possess conformal invariance, then one knows that time and space can be interchanged, $t \leftrightarrow x$ or equivalently, the system is invariant under the modular transformation $S$. One now has a cylinder which is wrapped around in the spatial direction and extending upward in time. The corresponding Hamiltonian in this picture can be regarded as propagating the system for the time interval $\ell$ from the initial and final state $\vert A\rangle,\vert B \rangle$
\begin{equation}
	Z_{AB} = \langle A \vert e^{-\ell H^{\beta} } \vert B \rangle \label{eqn:cstring} .
\end{equation}
The states here belong to the Hilbert space of states quantized on the circle, \emph{i.e.} they can be decomposed into linear combinations of states in the representation of the Virasoro algebra which are labelled by $(h,\bar{h})$, the highest weights. \cite{bigyellowbk}

Because conformal invariance is so restrictive in two dimensions, one can say more about the boundary states $\vert A\rangle$. One typically imposes the condition that $T(z) = \bar{T}(\bar{z})$ where $T,\bar{T}$ are the holomorphic and anti-holomorphic components of the stress energy tensor, $z=t+ix$. In the $x,t$ basis, this means that the diagonal components of the stress energy tensor vanish at the boundary $T_{x,t}$. If the boundary is in the time direction, this means no momentum flows out of the system. The stress energy generates the conformal symmetry so that the boundary states must satisfy the condition,
\begin{equation}
	\left[ T - \bar{T} \right] \vert A \rangle =0 \label{eqn:cfBC} .
\end{equation}
Fourier transforming, this can be written in terms of Virasoro generators,
\begin{equation}
	\left[ L_{n} -\bar{L}_{-n} \right] \vert A \rangle = 0 .
\end{equation}
This implies that the boundary state $\vert A \rangle$ must made out of states with the holomorphic and anti-holomorphic sectors stitched together in a specific way, \emph{i.e.}
\begin{equation}
	\vert h \rangle \rangle = \sum_{m} \vert h ; m \rangle \otimes \overline{\vert h ; -m \rangle} .
\end{equation}
Here, $m$ labels the descendant level in the representation $h$ that belongs to the subset of representations that appear simultaneously in the holomorphic and anti-holomorphic sectors of the Virasoro algebra. The state $\vert h \rangle\rangle$ are known as the Ishibashi states.\cite{ishibashi}

The Ishibashi states turn out to form a basis for the possible boundary states, and one can write an arbitrary state $\vert A \rangle$ as a linear combination of the Ishibashi states,
\begin{equation}
	\vert A \rangle = \sum_{i} C^{i}_{A} \vert i \rangle\rangle. \label{eqn:bstate} 
\end{equation}
Hence, the characterization of a conformal boundary condition is reduced to finding the matrix elements $C^{i}_{A}$. Now, using the expression (\ref{eqn:bstate}) into (\ref{eqn:cstring}) one finds that
\begin{equation}
	Z_{AB} = \sum_{i} C^{i}_{A} C^{B}_{i} \langle\langle i \vert e^{-\ell H^{\beta} } \vert i \rangle\rangle .
\end{equation}
The overlap can be identified with the character of the representation $i$, $\chi_{i}\left( e^{-4\pi \ell/\beta} \right)$. Now noting that (\ref{eqn:ostring}) can be written as a sum of characters, and using the fact that the two quantities are in fact equivalent by conformal invariance leads one to the relationship
\begin{equation}
	\sum_{i} C^{i}_{A} C^{B}_{i} \chi_{i} \left( e^{-4\pi \ell/\beta} \right) = \sum_{i} n^{i}_{AB} \chi_{i}\left( e^{-\pi \beta/\ell} \right) ,
\end{equation}
where $n^{i}_{AB}$ are the multiplicities that indicate the number of times a representation $i$ appears in the Hilbert space with boundary conditions $A$ and $B$. One then notes that the characters transform among themselves by the modular $S$-matrix so that
\begin{equation}
	 \chi_{i} \left( e^{-\pi \beta/\ell} \right) = \sum_{j} S_{i}^{j} \chi_{j} \left( e^{-4\pi \ell/\beta} \right) .
\end{equation}
If the characters are linearly independent, then this leads one to Cardy's equation which relates the multiplicities $n^{i}_{AB}$ that characterize the spectrum of the theory for fixed boundary conditions $A,B$ and the matrix elements $C^{i}_{A}$ that characterize the boundary states,
\begin{equation}
	\sum_{j} S^{j}_{i} n^{i}_{AB} = C^{i}_{A} C^{B}_{i}.
\end{equation}
The key problem in boundary conformal field theory is finding a set of boundary states where the multiplicities are non-negative integers.\cite{Cardy1989,Cardy1991} For the free boson, a solution to this requirement is
\begin{equation}
	C_{A}^{i} \propto \sum_{w,n=-\infty}^{\infty} e^{i \hat{\pi}_{0}(i) \varphi_{0} } \label{eqn:bosonBstates} ,
\end{equation}
where $\pi_{0}$ and $\varphi_{0}$ are defined in (\ref{eqn:zeromode}).

Note that linear combinations of boundary states $\vert A \rangle$ also satisfy the above constraints. An additional choice that is imposed is that $n^{0}_{AA} = 1$, that is to say that the identity representation appears exactly once in the spectrum of the theory with $A,A$ boundary conditions.   
Operatively, this fixes the normalization of the boundary states so that in the long cylinder limit $Z_{AA}$ contains the identity exactly once. This gives the $g$-factor of the boundary state,\cite{Affleck1991}
\begin{equation}
	g_{A} = \langle 0 \vert A\rangle	\label{eqn:gfactor}.
\end{equation}
It has been conjectured that relevant boundary perturbations drive the system to fixed points given by lower values of the $g$-factor.\cite{Affleck1991} In this sense, the $g$-factor is also a characteristic of the boundary condition.

\section{$\vartheta$-functions}
\label{sec:AppendixTheta}

The $\vartheta$-functions are defined as
\begin{eqnarray}
	\vartheta_{1}(\nu\vert\tau) &=& i \sum_{n=-\infty}^{\infty} (-1)^{n} q^{\frac{1}{2}(n-1/2)^{2} } e^{i\pi(2n-1)\nu} \label{eqn:theta1}  \\
	&=& 2 q^{1/8} \sin(\pi \nu) f(q)   \prod_{m=1} \left(1-2 \cos(2\pi \nu) q^{m} +q^{2m} \right)  ,\nonumber \\
	\vartheta_{2}(\nu\vert\tau) &=& \sum_{n=-\infty}^{\infty} q^{\frac{1}{2}(n-1/2)^{2}} e^{i\pi(2n-1)\nu} \label{eqn:theta2} \\ 
	&=& 2 q^{1/8} \cos(\pi\nu) f(q)\prod_{m=1}^{\infty} \left(1 + 2\cos(2\pi\nu) q^{m} + q^{2m} \right) , \nonumber \\
	\vartheta_{3}(\nu\vert\tau) &=& \sum_{n=-\infty}^{\infty} q^{\frac{1}{2} n^{2} } e^{i 2\pi n \nu}   \label{eqn:theta3} \\
	&=& f(q) \prod_{m=1}^{\infty}  \left(1+2\cos(2\pi \nu) q^{n-1/2} + q^{2n-1} \right) , \nonumber \\
	\vartheta_{4}(\nu\vert\tau) &=& \sum_{n=-\infty}^{\infty} (-1)^{n} q^{\frac{1}{2}n^{2} } e^{i2\pi n\nu} \label{eqn:theta4}\\
	&=& f(q) \prod_{m=1}^{\infty}\left(1-2\cos(2\pi\nu) q^{n-1/2} + q^{2n-1}\right) , \nonumber
\end{eqnarray}
where $q=e^{2\pi i \tau}$ and
\begin{equation}
	f(q) = \prod_{m=1}^{\infty} \left(1-q^{m} \right) = \left(\frac{1}{2\pi q^{1/4} } \frac{\partial \vartheta_{1}(\nu\vert\tau)}{\partial \nu} \big\vert_{\nu=0} \right)^{1/3} .
\end{equation}
The $\eta(\tau)$ function is then defined as
\begin{equation}
	\eta(\tau) = q^{1/24} f(q) .
	\label{eqn:eta}
\end{equation}
The action of the modular transformation $S:\tau\rightarrow -1/\tau$ on $\vartheta_{k}$-functions can be found by making use of the Possion resummation formula,
\begin{eqnarray}
	\sum_{n=-\infty}^{\infty} e^{-\pi n^{2} A  + 2n \pi A s} = \frac{1}{\sqrt{A} } e^{\pi A s^{2} } \sum_{m=-\infty}^{\infty} e^{-\pi A^{-1} m^{2} - 2i\pi ms} .\nonumber \\
	\label{eqn:poisson}
\end{eqnarray}


\section*{References}

\begin{thebibliography}{10}

\bibitem{sachdev-book}
Subir Sachdev.
\newblock {\em {Quantum Phase Transitions}}.
\newblock Cambridge University Press, Cambridge, UK, 2001.

\bibitem{amico-2008}
Luigi Amico, Rosario Fazio, Andreas Osterloh, and Vlatko Vedral.
\newblock {Entanglement in Many Body Systems}.
\newblock {\em Rev. Mod. Phys.}, 80:517, 2008.

\bibitem{Srednicki1993}
Mark Srednicki.
\newblock {Entropy and Area}.
\newblock {\em Phys. Rev. Lett.}, 71:666, 1993.

\bibitem{Bombelli1986}
Luca Bombelli, Rabinder~K. Koul, Joohan Lee, and Rafael~D. Sorkin.
\newblock Quantum source of entropy for black holes.
\newblock {\em Phys. Rev. D}, 34:373, 1986.

\bibitem{Callan1994}
Curtis~G. Callan and Frank Wilczek.
\newblock {On geometric entropy}.
\newblock {\em Phys. Lett. B}, 333:55, 1994.

\bibitem{Holzhey1994}
Christopher Holzhey, Finn Larsen, and Frank Wilczek.
\newblock {Geometric and renormalized entropy in conformal field theory}.
\newblock {\em Nucl. Phys. B}, 424:443, 1994.

\bibitem{Calabrese2004}
Pasquale Calabrese and John Cardy.
\newblock {Entanglement entropy and quantum field theory}.
\newblock {\em J. Stat. Mech. JSTAT}, 04:{P06002}, 2004.

\bibitem{Vidal2003}
Guifre Vidal, Jos\'e~L. Latorre, Enrique Rico, and Alexei Kitaev.
\newblock {Entanglement in Quantum Critical Phenomena}.
\newblock {\em Phys. Rev. Lett.}, 90:227902, 2003.

\bibitem{Hsu2009b}
Benjamin Hsu, Eytan Grosfeld, and Eduardo Fradkin.
\newblock Quantum noise generated by a quantum quench.
\newblock {\em Phys. Rev. B}, 80:235412, 2009.

\bibitem{Lehur2009}
H~Francis Song, Stephan Rachel, and Karyn LeHur.
\newblock {General Relation between Entanglement and Fluctuations in One
  Dimension }.
\newblock (unpublished) arXiv:1002.0825, 2010.

\bibitem{KlichLevitov}
Israel Klich and Leonid Levitov.
\newblock Quantum noise as an entanglement meter.
\newblock {\em Phys. Rev. Lett.}, 102(10):100502, 2009.

\bibitem{CalabreseCardy-CFT}
Pasquale Calabrese and John Cardy.
\newblock Entanglement and correlation functions following a local quench: a
  conformal field theory approach.
\newblock {\em J. Stat. Mech. JSTAT}, 2007(10):P10004, 2007.

\bibitem{refael-2009}
Gil Refael and Joel~E. Moore.
\newblock {Criticality and entanglement in random quantum systems}.
\newblock {\em J. Phys. A: Math. and Theor.}, 42:504010, 2009.

\bibitem{Chakravarty}
Sudip Chakravarty.
\newblock {Scaling of the von Neumann entropy at the Anderson transition}.
\newblock In Elihu Abrahams, editor, {\em {Fifty years of Anderson
  Localization}}, Singapore, 2010. World Scientific.
\newblock (in press), arXiv: 1004.0730v1.

\bibitem{Kitaev2006}
Alexei Kitaev and John Preskill.
\newblock {Topological Entanglement Entropy}.
\newblock {\em Phys. Rev. Lett.}, 96:110404, 2006.

\bibitem{Levin2006}
Michael Levin and Xiao~Gang Wen.
\newblock Detecting topological order in a ground state wave function.
\newblock {\em Phys. Rev. Lett.}, 96:110405, 2006.

\bibitem{Dong2008}
Shying Dong, Eduardo Fradkin, Robert~G. Leigh, and Sean Nowling.
\newblock {Topological Entanglement Entropy in Chern-Simons Theories and
  Quantum Hall Fluids}.
\newblock {\em J. High Energy Phys. JHEP}, 05:016, 2008.

\bibitem{Fradkin2006}
Eduardo Fradkin and Joel~E. Moore.
\newblock {Entanglement entropy of 2D conformal quantum critical points:
  hearing the shape of a quantum drum}.
\newblock {\em Phys. Rev. Lett.}, 97:050404, 2006.

\bibitem{Hsu2009}
Benjamin Hsu, Michael Mulligan, Eduardo Fradkin, and Eun-Ah Kim.
\newblock Universal entanglement entropy.
\newblock {\em Phys. Rev. B}, 79:115421, 2009.

\bibitem{pasquier2009}
Jean~Marie St{\'e}phan, Shunsuke Furukawa, Gregoire Misguich, and Vincent
  Pasquier.
\newblock Shannon and entanglement entropies of one and two dimensional
  critical wave functions.
\newblock {\em Phys. Rev. B}, 80(184421), 2009.

\bibitem{Metlitski}
Max~A Metlitski, Carlos~A Fuertes, and Subir Sachdev.
\newblock {Entanglement Entropy in the O(N) Models}.
\newblock {\em Phys. Rev. B}, 80:115122, 2009.

\bibitem{FradkinReview}
Eduardo Fradkin.
\newblock Scaling of the entanglement entropy.
\newblock {\em J. Phys. A: Math. Theor.}, 42:504011, 2009.

\bibitem{Ardonne2004}
Eddy Ardonne, Paul Fendley, and Eduardo Fradkin.
\newblock {Topological Order and Conformal Quantum Critical Points}.
\newblock {\em Ann. Phys.}, 310:493, 2004.

\bibitem{Rokhsar1988}
Daniel~S. Rokhsar and Steven~A. Kivelson.
\newblock {Superconductivity and the Quantum Hard-Core Dimer Gas}.
\newblock {\em Phys. Rev. Lett.}, 61:2376, 1988.

\bibitem{Castelnovo2005}
Claudio Castelnovo, Claudio Chamon, Christopher Mudry, and Pierre Pujol.
\newblock {From quantum mechanics to classical statistical physics: generalized
  Rokhsar-Kivelson Hamiltonians and the ``Stochastic Matrix Form''
  decomposition}.
\newblock {\em Ann. Phys.}, 318:316, 2005.

\bibitem{Fendley2006}
Paul Fendley.
\newblock Loop models and their critical points.
\newblock {\em J. Phys. A: Math. Theor.}, 39:15445, 2006.

\bibitem{Fendley2008}
Paul Fendley.
\newblock Topological order from quantum loops and nets.
\newblock {\em Ann. Phys.}, 323:3113, 2008.

\bibitem{Moessner2002}
Roderich Moessner, Shivaji~L. Sondhi, and Eduardo Fradkin.
\newblock {Short-ranged RVB physics, quantum dimer models and Ising gauge
  theories}.
\newblock {\em Phys. Rev. B}, 65:024504, 2002.

\bibitem{Henley1997}
Christopher~L. Henley.
\newblock {Relaxation time for a dimer covering with height representation}.
\newblock {\em J. Stat. Phys.}, 89:483, 1997.

\bibitem{Oshikawa1997}
Masaki Oshikawa and Ian Affleck.
\newblock {Boundary conformal field theory approach to the critical
  two-dimensional Ising model with a defect line}.
\newblock {\em Nucl. Phys. B.}, 495:533, 1997.

\bibitem{Kac1966}
Marc Kac.
\newblock Can you hear the shape of a drum?
\newblock {\em Amer. Math. Monthly}, 73:1, 1966.

\bibitem{Privman1984}
Vladimir Privman and Michael~E. Fisher.
\newblock {Universal critical amplitudes in finite-size scaling}.
\newblock {\em Phys. Rev. B}, 30:322, 1984.

\bibitem{CardyPeschel}
John~L. Cardy and Ingo Peschel.
\newblock Finite size dependence of the free energy in two dimensional critical
  systems.
\newblock {\em Nucl. Phys. B}, 300:377, 1988.

\bibitem{Privman1988}
Vladimir Privman.
\newblock {Universal Size Dependence of the Free Energy of Finite Systems Near
  Criticality}.
\newblock {\em Phys. Rev. B}, 38:9261--9263, 1988.

\bibitem{Cardy1989}
John~L. Cardy.
\newblock {Boundary conditions, fusion rules, and the Verlinde formula}.
\newblock {\em Nucl. Phys. B}, 324:581, 1989.

\bibitem{Furukawa2007}
Shunsuke Furukawa and Gregoire Misguich.
\newblock {Topological Entanglement Entropy in the Quantum Dimer Model on the
  Triangular Lattice}.
\newblock {\em Phys. Rev. B}, 75:214407, 2007.

\bibitem{affleck2001}
Ian Affleck, Masaki Oshikawa, and Hubert Saleur.
\newblock {Quantum Brownian Motion on a Triangular Lattice}.
\newblock {\em Nucl. Phys. B}, 594:535, 2001.

\bibitem{Affleck2008}
Ian Affleck.
\newblock {Quantum Impurity Problems in Condensed Matter Physics}.
\newblock In J.~Jacobsen, S.~Ouvry, V.~Pasquier, D.~Serban, and L.F.
  Cugliandolo, editors, {\em {Les Houches 2008, Session LXXXIX: Exact Methods
  in Low-Dimensional Statistical Physics and Quantum Computing}}, Oxford, UK,
  2010. Oxford University Press.

\bibitem{oshikawa2006}
Masaki Oshikawa, Claudio Chamon, and Ian Affleck.
\newblock Junctions of three quantum wires.
\newblock {\em J. Stat. Mech. JSTAT}, 06:P02008, 2006.

\bibitem{CaldeiraLeggett3}
Amir~O. Caldeira and Anthony~J. Leggett.
\newblock {Path Integral Approach to Quantum Brownian Motion}.
\newblock {\em Physica A}, 121:587, 1983.

\bibitem{hofstadter}
Douglas~R Hofstadter.
\newblock {Energy levels and wave functions of Bloch electrons in rational and
  irrational magnetic fields}.
\newblock {\em Phys. Rev. B}, 14:2239, 1976.

\bibitem{callan1995}
Curtis~G. Callan, Igor~R. Klebanov, Juan~M. Maldacena, and Ali Yegulalp.
\newblock {Magnetic Field and Fractional Statistics in BCFT}.
\newblock {\em Nucl. Phys. B}, 443:444, 1995.

\bibitem{CallanFreed}
Curtis~G. Callan and Denise Freed.
\newblock Phase diagram of the dissipative hofstadter model.
\newblock {\em Nucl. Phys. B}, 374:543, 1992.

\bibitem{Haldane2008}
Hui Li and F.~Duncan~M. Haldane.
\newblock {Entanglement Spectrum as a Generalization of Entanglement Entropy:
  Identification of Topological Order in Non-Abelian Fractional Quantum Hall
  Effect States}.
\newblock {\em Phys. Rev. Lett.}, 101:010504, 2008.

\bibitem{oshikawa}
Masaki Oshikawa.
\newblock Boundary conformal field theory and entanglement entropy in
  two-dimensional quantum lifshitz critical point.
\newblock arXiv: 1007:3739.

\bibitem{Alet2005}
Fabien Alet, Jesper~Lykke Jacobsen, Gregoire Misguich, Vincent Pasquier,
  Frederic Mila, and Matthias Troyer.
\newblock {Interacting Classical Dimers on the Square Lattice}.
\newblock {\em Phys. Rev. Lett.}, 94:235702, 2005.

\bibitem{Baxter1982}
Rodney~J. Baxter.
\newblock {\em {Exactly Solved Models in {S}tatistical {M}echanics}}.
\newblock Academic Press, 1982.
\newblock and references therein.

\bibitem{bigyellowbk}
Philippe~D. Francesco, Pierre Mathieu, and David S\'en\'echal.
\newblock {\em {Conformal Field Theory}}.
\newblock Springer-Verlag, 1997.

\bibitem{Weisberger87}
William~I. Weisberger.
\newblock {Conformal Invariants for Determinants of Laplacians on Riemann
  Surfaces}.
\newblock {\em Comm. Math. Phys.}, 112:633, 1987.

\bibitem{CardyCalabrese2010}
John~L. Cardy and Pasquale Calabrese.
\newblock Unusual corrections to scaling in entanglement entropy.
\newblock {\em J. Stat. Mech.}, P04023, 2010.

\bibitem{gswv1}
Michael~B. Green, John~H. Schwarz, and Edward Witten.
\newblock {\em Superstring Theory}, volume~1.
\newblock Cambridge Monographs on Mathematical Physics, 1987.

\bibitem{frohlich}
Jurg Fr\"ohlich, Jurgen Fuchs, Ingo Runkel, and Christopher Schweigert.
\newblock {Duality and Defects in RCFT}.
\newblock {\em Nucl. Phys. B}, 763:354, 2007.

\bibitem{fuchs97}
Jurgen Fuchs, Matthias~R Gaberdiel, Ingo Runkel, and Christopher Schweigert.
\newblock {Topological Defects for Free Boson CFT}.
\newblock {\em J. Phys. A: Math. and Theor.}, 40:11403, 2007.

\bibitem{Callan1986}
Ahmed Abouelsaood, Curtis~G. Callan, Chiara~R. Nappi, and Scott~A. Yost.
\newblock {Open Strings in Background Gauge Fields}.
\newblock {\em Nucl. Phys. B}, 280:599, 1987.

\bibitem{Freedman2004}
Michael Freedman, Chetan Nayak, Krill Shtengel, and Kevin Walker.
\newblock {A Class of ${P},{T}$-Invariant Topological Phases of Interacting
  Electrons}.
\newblock {\em Ann. Phys.}, 310:428, 2004.

\bibitem{Levin2005}
Michael Levin and Xiao-Gang Wen.
\newblock {String-net condensation: A physical mechanism for topological
  phases}.
\newblock {\em Phys. Rev. B}, 71:045110, 2005.

\bibitem{Hsu2010b}
Benjamin Hsu and Eduardo Fradkin.
\newblock Entanglement without pain.
\newblock (in preparation), 2010.

\bibitem{ishibashi}
Nobuyuki Ishibashi.
\newblock {The Boundary and Crosscap States in Conformal Field Theories }.
\newblock {\em Mod. Phys. Lett. A}, 4:251, 1989.

\bibitem{Cardy1991}
John~L. Cardy and David~C Lewellen.
\newblock Bulk and boundary correlation functions.
\newblock {\em Phys. Lett. B}, 259:274, 1991.

\bibitem{Affleck1991}
Ian Affleck and Andreas W.~W. Ludwig.
\newblock {Universal Non-Integer Groundstate Degeneracy}.
\newblock {\em Phys. Rev. Lett.}, 67:161, 1991.

\end{thebibliography}
\end{document}